\documentclass[article,accept,moreauthors,pdftex,10pt,a4paper,preprints]{mdpi} 

\firstpage{1} 

\history{}

\usepackage[utf8]{inputenc}
\usepackage[T1]{fontenc}
\usepackage{hyperref}
\usepackage{graphicx,bm,amsmath,color,scalerel}
\usepackage{bbold}
\usepackage{wasysym}
\usepackage{verbatim}
\usepackage{physics}
\usepackage{comment}
\usepackage{subcaption}
\graphicspath{ {./img/} }
\usepackage{nomencl}
\makenomenclature
\usepackage{etoolbox}
\usepackage{cancel}
\renewcommand\nomgroup[1]{%
  \item[\bfseries
  \ifstrequal{#1}{Z}{Abbreviations}{%
  \ifstrequal{#1}{N}{Number Sets}{%
  \ifstrequal{#1}{O}{Other Symbols}{}}}%
]}

\newcommand{\be}{\begin{equation}}
\newcommand{\ee}{\end{equation}}
\newcommand{\beq}{\begin{eqnarray}}
\newcommand{\eeq}{\end{eqnarray}}
\newcommand{\ba}{\begin{align}}
\newcommand{\ea}{\end{align}}
\newcommand{\red}[1]{\textcolor{red}{#1}}

\Title{Particle-antiparticle asymmetry in a relativistic deformed kinematics}

\Author{José Manuel Carmona$^{1}$*\orcidA{}, José Luis Cortés$^{1}$\orcidB{} and José Javier Relancio$^{2,3,4}$\orcidC{}}

\AuthorNames{José Manuel Carmona, José Luis Cortés and José Javier Relancio}

\address{%
$^{1}$ \quad Departamento de F\'{\i}sica Te\'orica and Centro de Astropartículas y Física de Altas Energías (CAPA),
Universidad de Zaragoza, Zaragoza 50009, Spain; jcarmona@unizar.es (J.M.C); cortes@unizar.es (J.L.C.);\\
$^{2}$  \quad Dipartimento di Fisica ``Ettore Pancini'', Università di Napoli Federico II, Napoli, Italy;\\
$^{3}$ \quad INFN, Sezione di Napoli, Italy; \\
$^{4}$ \quad Centro de Astropartículas y Física de Altas Energías (CAPA), Universidad de Zaragoza, Zaragoza 50009, Spain; relancio@unizar.es (J.J.R.)
}

\corres{Correspondence: jcarmona@unizar.es}

\abstract{Relativistic deformed kinematics are usually considered as a way to capture residual effects of a fundamental quantum gravity theory.  These kinematics present a non-commutative addition law for the momenta, so that the total momentum of a multi-particle system depends on the specific ordering in which the momenta are composed. We explore in the present work how this property may be used to generate an asymmetry between particles and antiparticles through a particular ordering prescription, resulting in a violation of \texorpdfstring{$CPT$}{CPT} symmetry. We study its consequences for muon decay, obtaining a difference in the lifetimes of the particle and the antiparticle as a function the new high-energy scale parameterizing such a relativistic deformed kinematics.}


\begin{document}

\medskip

\section{Introduction}
\label{sec:introduction}

The generation of a tiny (of the order of $10^{-9}$ ~\cite{Zyla:2020zbs}) baryonic asymmetry (and, more generally, an asymmetry between particles and antiparticles) in the early universe is one of the most important open problems in fundamental physics.

In his well-known work~\cite{Sakharov:1967dj}, Sakharov pointed out three basic conditions for the cosmological formation of a
baryonic asymmetry. The first one was the existence of processes that generate a net baryon number. The other two were established to guarantee that such net baryon number is maintained.

One of these additional requirements (the `third' Sakharov condition) asks for a departure from thermal equilibrium, since the equality of masses of particles and antiparticles (a consequence of $CPT$ symmetry) imply that their numbers have to be the same~\cite{Sakharov:1990gva}.

The remaining `second' Sakharov condition asks for the occurrence of a $C$ asymmetry (that is, an asymmetry between the behavior of particles and antiparticles) in the evolution of the universe. Indeed, there must be a difference between two charge-conjugated processes, that is, those related by changing all particles into antiparticles (and viceversa). If this were not the case, the net baryonic number generated in one of the processes would be compensated by the generation of an equal amount of net antibaryonic number in the second one.

Usually, the occurrence of such $C$ asymmetry is associated to the violation of $C$ and $CP$ symmetries. A violation of $C$ is not enough, since one could have a difference between the partial probabilities (the angular dependence of differential cross-sections) of the charge-conjugate reactions, reflecting a violation of $C$, while the total cross section could still be the same. In such a case, the combination of both processes would not generate a $C$ asymmetry. For example, a leptonic decay such as $\mu^-\to e^-+\bar{\nu}_e+\nu_\mu$ is not $C$-invariant, neither $P$-invariant, but the invariance under the combination of both operations, charge conjugation and parity, guarantees that the total decay rates of this process and its charge conjugated one are the same. A source of $CP$ violation in the dynamics of the interactions besides a possible violation of $C$ is then normally thought to be necessary to generate a $C$ asymmetry which could produce a difference in the number of particles and antiparticles. A $CP$ violation is indeed present in the standard model of particle physics, but in a quantity which is not large enough to account for the observed baryon asymmetry~\cite{Bambi:2015mba}.

In this paper we are going to explore another possibility for such a $C$ asymmetry, that is, that the two charge-conjugate processes have not the same probabilities because of a difference in the kinematics of these processes, instead of a difference coming from the dynamics. We will do that by introducing a new fundamental asymmetry between particles and antiparticles which reflects on the interactions where they participate. As a consequence, it will be possible to generate a matter-antimatter asymmetry by the substitution of the second and third Sakharov conditions by this new `kinematic asymmetry', which, as we will see, will imply a violation, not only of $C$ invariance, but also of $CPT$ invariance.

In order to introduce the kinematic asymmetry, one needs to go beyond the framework of the kinematics of special relativity (SR), where particles and antiparticles are treated on an equal footing. In fact, deviations from SR have been considered as a possible low-energy phenomenological footprint of a quantum theory of gravity~\cite{Amati:1988tn,Kostelecky1989,Garay1995,Amelino-Camelia1998,Gambini1999,Seiberg:1999vs,Alfaro:1999wd,Yoneya:2000bt,Jacobson:2002hd,Alfaro:2004aa,Collins:2004bp,Jacobson:2005bg,Hagar:2009zz,Horava:2009uw,Hossenfelder:2012jw}. This is seen as a natural hypothesis because of the special role that spacetime would have in a theory trying to incorporate the effects of the gravitational interaction at a quantum level~\cite{Oriti:2009zz}. 

Lorentz invariance is however a good symmetry of Nature at the energies we have explored so far, so that we will hypothesize that the kinematics of SR should be \emph{deformed} with corrections parameterized by a high-energy scale $\Lambda$ which would reflect the unknown ultraviolet physics, in such a way that the standard kinematics is recovered in the $\Lambda\to\infty$ limit. In the quantum gravity phenomenology scenario mentioned above, a natural guess for the value of $\Lambda$ could be the Planck mass, although we should not be guided by naturalness, since there are many possible scenarios in which the scale responsible for these corrections could have a difference of many orders of magnitude from the Planck scale~\cite{ArkaniHamed:1998rs,Ellis:2005qa,Horava:2009uw,Bojowald:2011hd,Carmona:2017oit,Oriti:2018tym,Freidel:2021wpl}.

It turns out that there are two different ways of considering a deformation of the kinematics of SR. A possibility is that Lorentz invariance is a low-energy approximate symmetry, which in fact is  broken, either fundamentally or in a spontaneous symmetry breaking scenario. We speak then of Lorentz invariance violation (LIV). In this framework, the modification at the kinematic level is a modification of the dispersion relations, where new terms that violate Lorentz symmetry are added to the dispersion relation of SR~\cite{Colladay:1998fq}. A difference in the dispersion relation of a particle and its antiparticle has indeed been explored in a LIV context to account for the observed matter-antimatter asymmetry~\cite{Dolgov:1981hv,Cohen:1987vi,Bertolami:1996cq,Carmona:2004xc,DiGrezia:2005eez,Dolgov:2009yk,Mavromatos:2018map}.

In addition to LIV theories, there is a different scenario in which the Lorentz symmetry is not violated but changed into a new symmetry. In contrast to the LIV case, this framework maintains a relativistic principle by a suitable deformation of the standard Lorentz transformations. 

A realization of this scenario are the celebrated doubly special relativity (DSR) theories~\cite{AmelinoCamelia:2008qg}, which were initially motivated by the idea of introducing a length (the Planck length) as an additional invariant in the theory besides the speed of light. There have been a number of proposed DSR theories with different properties, such as the presence of a maximum energy or momentum scale~\cite{AmelinoCamelia:2000mn,AmelinoCamelia:2000ge,Magueijo:2001cr,AmelinoCamelia:2002gv}.

Generally speaking, this possibility can be referred to as the case of a relativistic deformed kinematics (RDK). While a modification in the dispersion relation was the only kinematic signature in the case of a LIV, an essential feature of an RDK, apart from modified Lorentz transformations and a possible deformed dispersion relation, is a non-additive composition of momenta in the conservation laws of a process. The appearance of non-linear $\Lambda$ dependent terms in the composition of momenta is a necessary ingredient in an RDK required by the compatibility of the deformed kinematics with the relativity principle~\cite{Carmona:2012un}.

We will consider a non-commutative composition of momenta to ensure that the considered RDK is not simply the kinematics of SR rewritten in non-standard momentum variables~\cite{Carmona:2012un} (a non-commutative composition of momenta cannot be turned into a commutative one by a non-linear change of momentum variables). As we will see, this non-commutativity of the composition law will provide us with a natural way of introducing an asymmetry between particles and antiparticles. The purpose of the present paper is just to illustrate this possibility in a particular example, leaving the potential consequences for the problem of baryogenesis for further work.

We now explain the structure of the paper. In Section~\ref{sec:RDK}, we define what we understand by a relativistic deformed kinematics and frame the specific RDK that we are going to consider. In Section~\ref{sec:asymmetry}, we explain how the deformed composition law can be used to define an asymmetry between particles and antiparticles in a rather natural way; its implications with respect to the standard $C$, $P$ and $T$ operations and their combination are then considered in Section~\ref{sec:CPT}, where we conclude that such an asymmetry implies a violation of $CPT$ invariance. Observable consequences are then explored in Section~\ref{sec:effects}.

Ours is not the first attempt to introduce a particle-antiparticle asymmetry in a relativistic theory beyond special relativity. Other approaches have recently appeared in the literature~\cite{Arzano:2019toz,Arzano:2020rzu}, although with essential differences and disparate implications. A comparison with these works is done in Section~\ref{sec:comparison}.

We conclude and summarize our results in Section~\ref{sec:conclusions}. Details of the  computations are included in the Appendix.

\section{Relativistic deformed kinematics}
\label{sec:RDK}

The kinematics of a process is the set of relations which  determine the possible values of the momenta of the particles in the final state of an scattering of two particles (the decay of a particle) for a given initial state of the two particles (particle). Those relations are: a dispersion relation for each particle, which determines the energy of each particle from its momentum, and a momentum-energy conservation law due to the invariance under translations in space-time. 

We refer to a relativistic kinematics when one has a set of observers (or reference frames) connected by a six-parameter continuous group of transformations which describes the kinematics with the same set of relations. The kinematics of special relativity is an example of a relativistic kinematics where the transformations between observers are the Lorentz transformations acting linearly on the energy and momentum of each particle.  
By a relativistic deformed kinematics we understand a set of relations depending on a new energy scale $\Lambda$ such that, in the limit where the energies of all the particles are much smaller than the scale $\Lambda$, the relations defining the kinematics reduce to the relations of special relativity and the transformations between observers reduce to the linear Lorentz transformations. 

In the following, we will consider a deformed relativistic kinematics such that the dispersion relation, and then the relation between the energy and the momentum of each particle, does not depend on the new scale $\Lambda$. This choice will allow us to illustrate in the simplest way the idea that a relativistic deformation of the kinematics provides a new way to introduce a particle-antiparticle asymmetry. 

We will also restrict ourselves to a deformed relativistic kinematics where the rotations acting linearly on the momenta of the particles are transformations among observers with the same kinematics. All the effects of the deformation (dependence on the scale $\Lambda$) will be introduced through the replacement of the addition of four momenta, which defines the energy-momentum conservation of special relativity, by a deformed associative non-commutative composition law of four-momenta ($\oplus$)
\be
(p\oplus q)_\mu \,=\, p_\mu + q_\mu + ...\,, \quad\quad\quad
(p\oplus q)_\mu \neq (q\oplus p)_\mu\,.
\label{eq:DCL-nc}
\ee
The non-commutativity of the composition law guarantees that we are not simply considering the kinematics of special relativity in a peculiar choice of energy-momentum variables. The associativity reduces the arbitrariness in the explicit form of the deformed kinematics relations in terms of the deformed composition law $\oplus$.

The dots in Eq.~\eqref{eq:DCL-nc} refer to terms which can be neglected in the limit where all the energies are much smaller than the scale $\Lambda$ and such that the deformed composition law is compatible with the invariance under rotations acting linearly on the momenta. Compatibility with the invariance under boosts, however, requires that they act nonlinearly on at least one of the momenta (that can be taken, for example, as the one which is on the right in the composition law). The momentum of the other can be taken to transform linearly, as well as the total momentum~\cite{Carmona:2021gbg}. The associativity of the composition law allows one to generalize this property to the case of a composition of an arbitrary number of particles, so that the total momentum can always be taken to transform linearly. We will use this property in section~\ref{sec:effects}.

Examples of deformed composition laws ($\oplus$) can be obtained from an algebraic perspective through the relation between  a relativistic deformed kinematics and a $\kappa$-Poincaré Hopf algebra~\cite{Majid1994} or alternatively from the relation of a relativistic deformed kinematics and the geometry of a maximally symmetric curved momentum space~\cite{Lobo:2016blj,Carmona:2019fwf} (see also~\cite{AmelinoCamelia:2011bm,Barcaroli:2015xda,Barcaroli:2016yrl} for different geometrical interpretations of a DRK). 

In particular, an example of a deformed kinematics in which the dispersion relation is the one of SR and all the modification is introduced at the level of a deformed composition law is the classical basis of $\kappa$-Poincaré kinematics~\cite{Borowiec2010}.
Here, while at the level of one particle there is not any modification of the kinematics, the multi-particle sector shows a completely new behavior due to the presence of the deformed composition law. This implies that, at the phenomenological level, the crucial difference with the SR scenario should be observed in scattering processes, instead of considering only free propagating particles. 

The deformed composition law in this basis reads
\begin{equation}
    (p\oplus q)_0\,=\, p_0\, \Pi(q)+\Pi^{-1}(p)\left(q_0+\frac{\vec{p}\cdot \vec{q}}{\Lambda}\right)\,,\qquad    (p\oplus q)_i\,=\, p_i \,\Pi(q)+q_i\,,
    \label{eq:dcl_bc}
\end{equation}
being
\begin{equation}
   \Pi(k)\,=\, \frac{k_0}{\Lambda}+\sqrt{1+\frac{k_0^2-\vec{k}^2}{\Lambda^2}}\,,\qquad  \Pi^{-1}(k)\,=\, \left(\sqrt{1+\frac{k_0^2-\vec{k}^2}{\Lambda^2}}-\frac{k_0}{\Lambda}\right)\left(1-\frac{\vec{k}^2}{\Lambda^2}\right)^{-1}\,,
\end{equation}
and where $\Lambda$ plays the role of the high-energy scale deforming the kinematics.

In the language of Hopf algebras, a deformed composition law such as Eq.~\eqref{eq:dcl_bc} corresponds to a non-trivial form of the mathematical structure know as coproduct~\cite{Carmona:2016obd}. Although it is a conjecture that the low-energy limit of a quantum theory of gravity will include such mathematical construction, a modified composition law of energy and momentum is a necessary ingredient in any attempt to deform special relativity in a way compatible with relativistic invariance. The modified composition law that we are using in the present work is just one example where there is no modification of the energy-momentum relation and then the strong constrains on the energy scale of the deformation due to observable consequences of a modification of the energy-momentum relation do not apply. 

\section{Particle-antiparticle asymmetry in a relativistic deformed kinematics}
\label{sec:asymmetry}

In order to illustrate the details of how the non-commutativity of the deformed composition law $\oplus$, which defines the relativistic deformed kinematics, allows one to introduce a particle-antiparticle asymmetry, we consider the kinematics of the decay of a muon
\be
\mu(p) \: \to \: \nu_\mu(p') \,+\, e(q) \,+\, \bar{\nu}_e(q')\,.
\ee
This is a simple enough process not to involve irrelevant complications. Moreover, it is
an interesting process from a phenomenological perspective to look for possible observable effects of a deformed relativistic kinematics. 

The deformation of the composition of four-momenta can be associated with the locality of the interaction responsible for the decay. Hence, it is natural to express the energy-momentum conservation relation in terms of the composition of the momenta $(q, q')$ and in terms of the composition of the momenta $(p, \hat{p'})$, where $\hat{k}$ is the antipode of a four-momentum $k$, defined by 
\be
k \oplus \hat{k} \,=\, \hat{k}\oplus k \,=\, 0\,.
\ee
We now introduce the particle-antiparticle asymmetry in the deformed relativistic kinematics by selecting the composition $(q\oplus q')$ of the momenta $(q, q')$. In this way, the total momentum of a particle-antiparticle system is defined as the composition of the momentum of the particle (which appears on the left of the composition law) with the momentum of the antiparticle (which appears to the right). This rule leads to consider two alternatives for the deformed energy-momentum conservation law
\be
1) \quad (p\oplus \hat{p'})_\mu \,=\, (q\oplus q')_\mu\,, \hskip 2cm 2) \quad (\hat{p'}\oplus p)_\mu \,=\, (q\oplus q')_\mu\,.
\ee
The associativity of the composition law allows to rewrite the deformed energy-momentum conservation as
\be
1) \quad p_\mu \,=\, (q\oplus q'\oplus p')_\mu\,, \hskip 2cm 2) \quad p_\mu \,=\, (p'\oplus q\oplus q')_\mu\,.
\ee
The first choice has on the right hand side the composition of the momentum of an antiparticle (the momentum $q'$ of the electron antineutrino) with the momentum of a particle (the momentum $p'$ of the muon neutrino). This goes against the rule selected for the total momentum as composing the momenta of particles with the momenta of antiparticles and not the other way round. The second alternative, however, leads to an expression for the total energy-momentum of the final state with a definite ordering of the momenta, with the momenta of particles appearing to the left of the momenta of antiparticles.
We then define the relativistic deformed kinematics of the decay of the muon by the energy-momentum conservation relation
\be
p_\mu \,=\, (p'\oplus q\oplus q')_\mu\,,
\label{DCLmu-}
\ee
together with the energy-momentum relation of special relativity for each of the four (anti)particles. 

If we repeat the same arguments for the decay of a $\mu^+$
\be
\mu^+(p) \: \to \: \bar{\nu}_\mu(p') \,+\, e^+(q) \,+\, \nu_e(q')\,,
\ee
we would select the composition $(q'\oplus q)$ of the momenta $(q,q')$, and the two alternatives for the deformed energy-momentum conservation would be 
\begin{align}
1) \quad (p\oplus \hat{p'})_\mu \,=\, (q'\oplus q)_\mu \quad \Rightarrow \quad p_\mu \,=\, (q'\oplus q\oplus p')_\mu\,, \nonumber \\
2) \quad (\hat{p'}\oplus p)_\mu \,=\, (q'\oplus q)_\mu \quad \Rightarrow \quad p_\mu \,=\, (p'\oplus q'\oplus q)_\mu\,.
\end{align}
In this case, the alternative which involves a definite ordering of the four-momenta of particles and antiparticles in the expression of the total four-momentum of the final state is the first one. Then, it defines the relativistic deformed kinematics of the decay of $\mu^+$ by the energy-momentum conservation relation 
\be
p_\mu \,=\, (q'\oplus q\oplus p')_\mu\,,
\label{DCLmu+}
\ee
which differs from the energy-momentum conservation relation (\ref{DCLmu-}) of the decay of a $\mu^-$. 

In summary, we introduce a particle-antiparticle asymmetry at the level of the deformed composition of four-momenta, by postulating a definite order in which the momenta of particles and antiparticles appear in the non-trivial four-momentum conservation law in the interaction vertex. This means that specific and different conservation laws apply to charge-conjugated processes among the different possible ``channels'', corresponding to permutations of momenta in the composition law, which will result in a different kinematics for the decay of a particle and its antiparticle.

Our proposal based on the introduction of a definite ordering of the momenta of particles and antiparticles implicitly assumes that an implementation of the relativistic deformed kinematics does not need to treat all channels on an equal footing. Such a proposal lies outside of previous attempts to formulate an interacting quantum field theory on the $\kappa$-deformed momentum space in the path-integral formalism, in which each of the possible channels contribute equally to the amplitude of a given process~\cite{AmelinoCamelia:2001fd,Freidel:2013rra}.

\section{Discrete symmetries}
\label{sec:CPT}

The particle-antiparticle asymmetry introduced through the non-commutativity of the composition of four-momenta in a relativistic deformed kinematics leads to study the relation between the deformation of the kinematics and the discrete transformations of parity ($P$), charge conjugation ($C$) and time reversal ($T$). 

By a parity transformation ($P$), one should compare the kinematics of a process and the kinematics of the process where one changes the sign of the momenta of all the particles. For the deformations we have considered in this work, the invariance under parity of the relativistic deformed kinematics is related to the absence of the third term in the general expression (compatible with the invariance under rotations acting linearly on the momenta of the particles) of the spacial components of the composition of two four-momenta
\be
(p\oplus q)_i \,=\, f \, p_i \, + \, g \, q_i \, + \, h \, \epsilon_{ijk} p_j q_k\,, 
\ee
where $f$, $g$, $h$ are functions depending on the four momenta $(p, q)$ and the scale $\Lambda$, which are invariant under rotations. Since these functions can depend only on $p_0/\Lambda$, $q_0/\Lambda$, $\vec{p}^2/\Lambda^2$, $\vec{q}^2/\Lambda^2$ and $\vec{p}\cdot\vec{q}/\Lambda^2$, they are invariant under the action of $P$. However, the symmetry under parity of the kinematics is lost if $h$ is nonzero. All the examples of relativistic deformed kinematics derived in the algebraic approach (based on Hopf algebras) or in the geometric approach (based on the geometry of a maximally symmetric curved momentum space) have a composition of momenta with $h=0$ and then the invariance under $P$ of the kinematics is not affected by the deformation.

By time reversal ($T$), one should compare the kinematics of a process and the kinematics of a process where one exchanges the initial and final states and changes the sign of the momenta of all the particles. The combination ($PT$) of time reversal and parity is just the exchange of initial and final states, something that has no effect on the energy-momentum conservation law and then the invariance under $PT$ of the kinematics is not affected by the deformation. Since the invariance under $P$ was not affected either, we conclude that time reversal is unaffected. 

By charge conjugation ($C$), one exchanges particles (antiparticles) by antiparticles (particles). The introduction of a particle-antiparticle asymmetry in the deformation of the relativistic kinematics leads then to a violation of the invariance under $C$ of the kinematics. 

We conclude that a relativistic deformed kinematics with a particle-antiparticle asymmetry automatically involves a violation of $CPT$. This is not surprising since a deformed composition law of four-momenta requires to go beyond local relativistic quantum field theory, which is the framework where the $CPT$-theorem applies. 

\section{Search for observable effects}
\label{sec:effects}

In order to look for possible observable consequences of the difference in the kinematics of the decay of a particle and its antiparticle when one has a relativistic deformed kinematics, we will compare their lifetimes or decay widths. 
For the purpose of this work, which is the determination of the term proportional to $(1/\Lambda)$ in the ratio of lifetimes of particles and antiparticles, one can neglect the dependence on the electron mass, since $m_e \ll m_\mu$.  

The decay width of $\mu^\pm$ in the case of SR kinematics is
\be
\Gamma \,=\, \frac{1}{2 E_p} \, \int  d\Phi_3(p;p',q,q')\: |{\cal M}|^2\,,
\ee
where one has a normalization factor ($1/2E_p$) from the initial state, an integral over the phase space of the final state of three particles
\begin{align}
d\Phi_3(p;p',q,q') \,=\, & \frac{d^4 p'}{(2\pi)^3} \delta(p'^2) \,\theta(p'_0) \, \frac{d^4 q}{(2\pi)^3} \delta(q^2) \, \theta(q_0)
 \, \frac{d^4 q'}{(2\pi)^3} \delta(q'^2) \, \theta(q'_0) \nonumber \\ & \: (2\pi)^4 \delta^4(p-p'-q-q')\,,
\end{align}
and the squared dynamical matrix element
\be
|{\cal M}|^2 \,=\, 64 \, G_F^2 (p\cdot q') (p'\cdot q)\,.
\ee

In the case of a relativistic deformed kinematics with no modification of the dispersion relation, one has for the $\mu^-$ decay width
\be
\widetilde{\Gamma}^- \,=\, \frac{1}{2 E_p} \, \int  d\widetilde{\Phi}^-_3(p;p',q,q')\: |\widetilde{{\cal M}}|^2\,.
\ee

The deformed phase space integral is identified from the condition that the successive composition of momenta transform as the momentum of one particle under the deformed Lorentz transformations 
\begin{align}
d\widetilde{\Phi}^-_3(p;p',q,q') \,=\, & \frac{d^4 p'}{(2\pi)^3} \delta(p'^2) \,\theta(p'_0) \,
\frac{d^4(p'\oplus q)}{(2\pi)^3} \,\delta(q^2) \, \theta(q_0)\, \frac{d^4(p'\oplus q\oplus q')}{(2\pi)^3} \delta(q'^2) \, \theta(q'_0) \nonumber \\ & \: (2\pi)^4 \delta^4[p-(p'\oplus q\oplus q')]\,.  
\end{align}
This is the simplest phase-space integral compatible with the Lorentz transformations, which act linearly over the variables in the integral (adequately chosen from the composition of momenta under consideration). It is possible to take a more general expression for the integral measure, based on a geometric interpretation~\cite{AmelinoCamelia:1999pm}, giving rise to an extra factor; in any case, this will disappear in the quotient of lifetimes that we will get at the end of the section.

We do not have a dynamical framework (deformed quantum field theory) to derive the deformed squared matrix element. All we can do is to use the compatibility with the relativistic invariance to try 
to identify the deformed squared matrix element. Using the relations
\begin{align}
p\cdot q' = \frac{1}{2} \left(p^2 +q'^2 - (p-q')^2\right) = \frac{1}{2} \left(m_\mu^2 - (p'+q)^2\right) \nonumber \\
p'\cdot q = \frac{1}{2} \left((p'+q)^2 - p'^2 - q^2\right) = \frac{1}{2} (p'+q)^2 \,,
\end{align}
the SR squared matrix element can be written as
\be
|{\cal M}|^2 \,=\, 16 \, G_F^2 \left(m_\mu^2 - (p'+q)^2\right) (p'+q)^2\,.
\ee

We define $|\widetilde{{\cal M}}|^2$ by the replacement  $(p'+q)^2 \to (p'\oplus q)^2$ for the $\mu^-$ decay, and  $(p'+q)^2 \to (q\oplus p')^2$ for the $\mu^+$ case. This choice comes from the way the variables $p'$ and $q$ appear in the energy-momentum conservation relations (\ref{DCLmu-}) and (\ref{DCLmu+}). 
In fact (see~\cite{Carmona:2021gbg}), the  nonlinear implementation of the Lorentz transformations on the momenta is fixed by the condition that both the composition law between momenta and the dispersion relation of the momenta which are being composed (which, in this case, is simply their square) have to be invariant (relativity principle). This means that the square of the composition of momenta $(p'\oplus q)$ in the $\mu^-$ decay, and $(q\oplus p')$ in the $\mu^+$ case, are invariant under Lorentz transformations.
Therefore, the deformed matrix elements defined in this way are the simplest choice of a deformation of the SR matrix element compatible with the deformed relativistic invariance.

Introducing the variable 
\be
k \doteq (p'\oplus q)\,,
\label{k-}
\ee
one has 
\be
q \,=\, (\widehat{p'}\oplus k)\,, \quad
q' \,=\, (\hat{k}\oplus p)\,,
\ee
and then
\be
\widetilde{\Gamma}^-\, =\,  \frac{G_F^2}{2\,\pi^5} \, \frac{1}{2 E_p} \int \frac{d^3 p'}{2 E_{p'}} \, d^4 k \, \delta[(\widehat{p'}\oplus k)^2] \, \theta[(\widehat{p'}\oplus k)_0] \, \delta[(\hat{k}\oplus p)^2] \, \theta[(\hat{k}\oplus p)_0] \, (m_\mu^2-k^2) \, k^2\,.
\label{Gamma-}
\ee
Appendix~\ref{appendix:decay} contains the details of the derivation of the final result in the reference frame in which the muon is at rest. We find that, at first order in an expansion in powers of  $(1/\Lambda)$, there is no change with respect the SR result, so we need to go to second order, 
\be
\widetilde{\Gamma}^- \,= \, \frac{G_F^2 m_\mu^5}{192\,\pi^3}\left(1-\frac{9\, m_\mu^2}{20\,\Lambda^2}\right)\,.
\label{Gamma-f}
\ee

In the case of the $\mu^+$ decay, one has the energy-momentum conservation relation (\ref{DCLmu+}) instead of (\ref{DCLmu-}). One introduces in this case the variable 
\be
k \,\doteq\, (q'\oplus q)
\ee
instead of (\ref{k-}) and then one has
\be
q \,=\, ( \widehat{q'}\oplus k), \quad
p'\,=\, (\hat{k}\oplus p).
\ee
The $\mu^+$ decay width is given by 
\be
\widetilde{\Gamma}^+\, =\,  \frac{G_F^2}{2\,\pi^5} \, \frac{1}{2 E_p} \int \frac{d^3 q'}{2 E_{q'}} \, d^4 k \, \delta[(\widehat{q'}\oplus k)^2] \, \theta[(\widehat{q'}\oplus k)_0] \, \delta[(\hat{k}\oplus p)^2] \, \theta[(\hat{k}\oplus p)_0] \, (m_\mu^2-(\hat{q}'\oplus p)^2) \, (\hat{q}'\oplus p)^2\,
\label{Gamma+}
\ee
instead of (\ref{Gamma-}) for the $\mu^-$ decay width. The phase space integral in the expressions for the decay widths of $\mu^-$ and $\mu^+$ are identical but the deformed matrix elements differ. In the $\mu^-$ decay matrix element one has the square of the variable of integration $k$ while in the $\mu^+$ case one has the square of the composition of the antipode of the variable of integration $q'$ and the momentum $p$ of the $\mu^+$. Indeed, we find (see Appendix~\ref{appendix:decay}) a different result for the $\mu^+$ decay width:
\be
\widetilde{\Gamma}^+ \,=\,  \frac{G_F^2 m_\mu^5}{192\,\pi^3}\left(1-\frac{ m_\mu^2}{2\,\Lambda^2}\right)\,.
\label{Gamma+f}
\ee

From (\ref{Gamma-f}) and (\ref{Gamma+f}), one has 
\be
\frac{\tilde{\tau}^+}{\tilde{\tau}^-} \,=\, \frac{\widetilde{\Gamma}^-}{\widetilde{\Gamma}^+} \,=\, \left(1 + \frac{m_\mu^2}{20\,\Lambda^2}\right)
\ee
for the ratio of lifetimes of $\mu^+$ and $\mu^-$, to be compared with the experimental result~\cite{Zyla:2020zbs}
\be
\left(\frac{\tau^+}{\tau^-}\right)_\text{exp} \,=\, 1 + (2\pm 8) \,10^{-5}\,.
\ee
From this comparison, we get a lower bound on the scale $\Lambda$ of RDK 
\be
\Lambda \,>\, 2.4 \:\text{GeV}\,.
\label{bound}
\ee
Should we had chosen the other ordering option (where the total momentum of a particle-antiparticle system is the composition of the momentum of the antiparticle with the momentum of the particle), the results of the $\mu^-$ and $\mu^+$ decay widths would have been interchanged. The lower bound on the scale $\Lambda$ of RDK would be in this case
\be
\Lambda \,>\, 3.1 \:\text{GeV}\,.
\label{bound2}
\ee

Usually, the scale of a deformation of the kinematics of SR is taken of the order of the Planck scale which is far above the bound (\ref{bound}). But one should take into account that all the bounds on the scale of departures from the SR kinematics which are of the order of the Planck scale are obtained assuming a modification of the dispersion relation. Therefore, they do not apply in the proposal of relativistic deformed kinematics considered in this work, which does not modify the dispersion relation. Another argument for such a choice for the typical scale is the possible origin of the deformation  of the kinematics in the quantum fluctuations of spacetime in a theory of quantum gravity. However, 
as we have mentioned in the introduction, there are many scenarios where the fundamental scale of such hypothetical theory is many orders of magnitude below the (Planck) scale of the classical theory of gravity.

A relativistic deformed kinematics which does not modify the dispersion relation will not affect the masses of particles; then, one does not get any bound on $\Lambda$ from the different limits on the difference of masses between a particle and its antiparticle. Other tests of $CPT$ involve hadrons which are not elementary particles and there is no clear way to give a theoretical prediction of the possible effects due to a relativistic deformation of the kinematics. 

There are other possible observable effects of a relativistic deformed kinematics beyond those involving a violation of $CPT$ symmetry which may well give more stringent bounds on $\Lambda$ than the bound (\ref{bound}) obtained from the lifetime measurements of $\mu^-$ and $\mu^+$. We have used the muon decay and its $C$-conjugated process as a very simple example allowing us to obtain a well-defined result for an observable effect, independently of its phenomenological relevance. 

\section{Comparison with previous approaches}
\label{sec:comparison}

In Refs.~\cite{Arzano:2019toz,Arzano:2020rzu}, a different way in which a $CPT$  violation could be present in the context of a RDK was considered, and, in particular, also in the classical basis of $\kappa$-Poincaré. In these works, it was assumed that particles and antiparticles satisfy the same dispersion relation, but the description of the momentum states of the antiparticles are depicted by the antipode of the composition law. Due to this takeover, different Lorentz transformations of particles and antiparticles were obtained, leading to different lifetimes when they do not decay at rest. In fact, in~\cite{Arzano:2019toz} it was claimed: \emph{It should be noted that the deformation of CPT symmetry we propose here leads to a subtle violation of Lorentz symmetry}. This goes against the relativity principle present in DSR theories, in which Lorentz symmetry is deformed but not violated.

Indeed, one finds some peculiarities in the proposed scenario. Particle and antiparticle which are both at rest for an observer, have different momenta for a boosted reference frame. Therefore, accepting the usual relationship between velocity and energy and momentum, i.e. $\vec{v}=\vec{p}/E$, it means that the velocity of the antiparticle differs from the one of the particle under a boost. This goes against the physical intuition of a relativistic theory for which both particle and antiparticle should move at the same velocity for the boosted observer.

The necessity to have a different action of boosts on particles and antiparticles appears as inevitable in a geometric context to try to maintain some form of Lorentz invariance~\cite{Arzano:2009ci}.
However, from a pure kinematic perspective, and avoiding any interpretation on the geometry of momentum space, it is completely plausible to consider that both particles and antiparticles are subject to the same linear Lorentz transformations. This is the most natural scenario which keeps the relativity principle. Then, with this consideration, the conclusion that a relativistic deformed kinematics leads to a difference of lifetimes for a particle and its antiparticle due to a difference in the Lorentz transformation, can be avoided.

Instead of considering the consequences of a possible $CPT$ violation in a RDK regarding the symmetries  of the one-particle system, in this work we present a model in which this violation arises when taking into account a multi-particle system involving the DCL, which is the main ingredient of a RDK. Moreover, we obtain that the difference in lifetimes arises when considering the decay at rest and then, since we are considering a deformed kinematics satisfying the relativity principle, this difference will exist in any reference frame.  

In summary, the two proposals have completely different sources of $CPT$ violation: either to consider that antiparticles are described by antipodes in the case of Refs.~\cite{Arzano:2019toz,Arzano:2020rzu}, or the prescription in the momenta ordering for particles and antiparticles in the present work.
Due to the difference in the two approaches, there is also a big difference in the constraint one obtains on the high-energy scale. While in~\cite{Arzano:2019toz,Arzano:2020rzu} it was obtained that $\Lambda>4\times10^{11}$\,TeV, here we obtain a result which is fourteen orders of magnitude lower. A large softening of the constraints in particle processes with respect to the Lorentz invariance violation case is in fact a feature of deformed theories that maintain a relativity principle~\cite{Carmona:2018xwm}.

\section{Conclusions}
\label{sec:conclusions}

In this work we have shown how an RDK allows us to introduce a new way to consider an asymmetry between particle and antiparticle at the kinematic level. The assumption  that the total momentum of a particle-antiparticle system is the composition (with a non-commutative deformed composition law) of the momentum of the particle and the momentum of the antiparticle leads to a violation of $CPT$. One consequence of this violation is a difference in the lifetimes of a particle and its antiparticle, which in the particular example explored in this work, muon decay, is proportional to the inverse of the square of the energy scale of the deformed kinematics. A comparison of the results with the experimental bounds on the ratio of lifetimes of muon and antimuon has led us to get a constraint on the high-energy scale of the order of GeV.

This work opens the window to a completely new phenomenology related to an asymmetry between particles and antiparticles in the kinematics. The compatibility of the deformed transformations between different observables with the dynamics can be used to identify the simplest possible deformation of the dynamical matrix elements consistent with relativistic invariance. This offers the possibility to explore possible observable effects of
a relativistic deformed kinematics in different processes of high-energy particle physics or, alternatively, to derive bounds on the associated new energy scale.

Another important conclusion of this work is that time delays may not necessarily be the best phenomenological window to DRK theories, as it is usually assumed for doubly special relativity scenarios, since there may be examples which show no effect on time delays (such as in the case where the dispersion relation of special relativity is not modified), but that can have (depending on the value of the high-energy scale) observable consequences in particle physics experiments. 

\authorcontributions{All authors contributed equally to the present work.}

\acknowledgments{We thank Michele Arzano and Jerzy Kowalski-Glikman for their useful feedback on the manuscript. This work is supported by Spanish grants PGC2018-095328-B-I00 (FEDER/Agencia estatal de investigación), and DGIID-DGA No. 2015-E24/2. JJR acknowledges support from the INFN Iniziativa Specifica GeoSymQFT. The authors would like to acknowledge the contribution of the COST Action CA18108 ``Quantum gravity phenomenology in the multi-messenger approach''.}

\conflictsofinterest{The authors declare no conflict of interest.} 





\appendixtitles{yes} 
\appendixsections{one} 
\appendix

\section{Computation of the decay width} 
\label{appendix:decay}

In this appendix we explicitly show the computations of the decay widths. We will consider the reference frame in which the muon is at rest; then, the four-momentum of the muon is $p=(m_\mu,0,0,0)$. Along the computation we will keep only first and second order contributions in an expansion in powers of the inverse of the high-energy scale $\Lambda$. Then we need only the first and second order terms of the composition law of two momenta $(l, r)$~\eqref{eq:dcl_bc} 
\begin{equation}
    (l\oplus r)_0\,=\, l_0+ r_0+\frac{\vec{l}\cdot \vec{r}}{\Lambda}+\frac{l_0 r^2-r_0 l^2+2 l_0 (l\cdot r) }{2\Lambda}\,,\qquad    (l\oplus r)_i\,=\, l_i\left(1+\frac{r_0}{\Lambda}+\frac{r^2}{2\Lambda}\right)+r_i\,,
    \label{eq:dcl_bc_1}
\end{equation}
and of the antipode of a momentum $l$
\begin{equation}
   \hat{l}_0\,=\, -l_0\left(1+\frac{\vec{l}^2}{\Lambda^2}\right)+\frac{\vec{l}^2}{\Lambda}\,,\qquad     \hat{l}_i\,=\, -l_i \left(1-\frac{l_0}{\Lambda}+\frac{l_0^2+\vec{l}^2}{2\Lambda}\right) \,.
    \label{eq:antipode}
\end{equation} 

The modification due to the deformed composition law in both decays requires to consider the composition of the antipode of a momentum with another momentum. From (\ref{eq:dcl_bc_1}) and (\ref{eq:antipode}, one finds
\be 
(\hat{l}\oplus r)^2 \,=\, (r-l)^2+\frac{(r^2-l\cdot r)(l^2-l\cdot r)}{\Lambda^2}\,.
\ee
This shows that there are no corrections at first order to the decay widths of $\mu^-$ and $\mu^+$. Then, we need to go to second order in the computations to see a difference with respect to the SR result.

One can use the Dirac delta functions to reduce the expression of the decay width to a double integral over an energy variable $E$ (the energy $E_{p'}$ of the $\nu_\mu$ in the decay of a $\mu^-$, or the energy $E_{q'}$ of the $\nu_e$ in the decay of $\mu^+$) and over an angular variable. Taking into account that the interval of the integration over the energy variable is $0 \leq E \leq (m_\mu/2 - m_\mu^3/8\Lambda^2)$, we obtain the results of the decay widths~\eqref{Gamma-f} and~\eqref{Gamma+f}.

\end{document}